%% file: manuscript.tex
\documentclass[10pt,a4paper]{article}
\usepackage[english]{babel}
\usepackage{textcomp}
\usepackage{mathpazo}
\usepackage{a4wide}
\usepackage{wrapfig}
\usepackage{graphicx}
\usepackage{amsmath,amssymb}
\usepackage{xcolor}
\definecolor{darkred}{rgb}{0.5,0,0}
\definecolor{darkblue}{rgb}{0,0,0.5}
\definecolor{firebrick}{rgb}{0.75,0.125,0.125}
\definecolor{darkgreen}{rgb}{0,0.5,0}
\usepackage[colorlinks=true,linkcolor=firebrick,citecolor=darkgreen,urlcolor=darkblue]{hyperref}
\usepackage{makecell}
\usepackage{booktabs}
\usepackage{enumitem}
\usepackage{multirow}
\usepackage{textpos}
\usepackage{cite}

\graphicspath{{figs/}}
\DeclareGraphicsExtensions{.pdf,.png,.jpg}

\def\avg#1{\langle{#1}\rangle}

\title{\textbf{Observation of a Large-scale Anisotropy in the Arrival
Directions of Cosmic Rays above 8\texttimes10\textsuperscript{18}\,eV}}

\author{The Pierre Auger Collaboration%
\footnote{correspondence to: \href{mailto:auger_spokespersons@fnal.gov}{auger\_spokespersons@fnal.gov}}
\footnote{The authors with their affiliations appear at the end of this article.}}

\date{22 September 2017}

\begin{document}

\maketitle

\begin{textblock}{7}(0,-4)
\noindent Published in Science as
\href{http://dx.doi.org/10.1126/science.aan4338}{DOI:\,10.1126/science.aan4338}
\end{textblock}

\begin{abstract}
Cosmic rays are atomic nuclei arriving from outer space that reach the highest
energies observed in nature. Clues to their origin come from studying the
distribution of their arrival directions. Using $3{\times}10^4$ cosmic rays
above $8{\times}10^{18}$\,electron volts, recorded with the Pierre Auger
Observatory from a total exposure of 76,800 square kilometers steradian year,
we report an anisotropy in the arrival directions.  The anisotropy, detected at
more than the 5.2$\sigma$ level of significance, can be described by a dipole
with an amplitude of $6.5_{-0.9}^{+1.3}$\% towards right ascension
$\alpha_\text{d}=100\pm10$\,degrees and declination
$\delta_\text{d}={-24_{-13}^{+12}}$\,degrees. That direction indicates an
extragalactic origin for these ultra-high energy particles.
\end{abstract}

Particles with energies ranging from below $10^9$\,eV up to beyond
$10^{20}$\,eV, known as cosmic rays, constantly hit the Earth's atmosphere. The
flux of these particles steeply decreases as their energy increases; for
energies above 10\,EeV ($1\,\text{EeV}\equiv10^{18}$\,eV), the flux is about
one particle per km$^2$ per year. The existence of cosmic rays with such
ultra-high energies has been known for more than 50 years~\cite{1,2}, but the
sites and mechanisms of their production remain a mystery. Information about
their origin can be obtained from the study of the energy spectrum and the mass
composition of cosmic rays. However, the most direct evidence of the location
of the progenitors is expected to come from studies of the distribution of
their arrival directions. Indications of possible hot spots in arrival
directions for cosmic rays with energy above 50\,EeV have been reported by the
Pierre Auger and Telescope Array Collaborations~\cite{3,4}, but the statistical
significance of these results is low. We report the observation, significant at
a level of more than 5.2$\sigma$, of a large-scale anisotropy in arrival
directions of cosmic rays above 8\,EeV.

Above $10^{14}$\,eV, cosmic rays entering the atmosphere create cascades of
particles (called extensive air-showers) that are sufficiently large to reach
the ground. At 10\,EeV, an extensive air-shower (hereafter shower) contains
${\sim}10^{10}$ particles spread over an area of ${\sim}20$\,km$^2$ in a thin
disc moving close to the speed of light. The showers contain an electromagnetic
component (electrons, positrons and photons) and a muonic component that can be
sampled using arrays of particle detectors. Charged particles in the shower
also excite nitrogen molecules in the air, producing fluorescence light that
can be observed with telescopes during clear nights.

The Pierre Auger Observatory, located near the city of Malarg\"ue, Argentina,
at latitude $35.2^\circ$S, is designed to detect showers produced by primary
cosmic rays above 0.1\,EeV. It is a hybrid system, a combination of an array of
particle detectors and a set of telescopes used to detect the fluorescence
light. Our analysis is based on data gathered from 1600 water-Cherenkov
detectors deployed over an area of 3000\,km$^2$ on a hexagonal grid with 1500-m
spacing.  Each detector contains 12\,tonnes of ultrapure water in a cylindrical
container, 1.2\,m deep and 10\,m$^2$ in area, viewed by three 9-inch
photomultipliers. A full description of the observatory, together with details
of the methods used to reconstruct the arrival directions and energies of
events, has been published~\cite{5}.

It is difficult to locate the sources of cosmic rays, as they are charged
particles and thus interact with the magnetic fields in our galaxy and the
intergalactic medium that lies between the sources and Earth. They undergo
angular deflections with amplitude proportional to their atomic number $Z$, to
the integral along the trajectory of the magnetic field (orthogonal to the
direction of propagation), and to the inverse of their energy $E$. At
$E\approx10$\,EeV, the best estimates for the mass of the particles~\cite{6}
lead to a mean value for $Z$ between 1.7 and 5. The exact number derived is
dependent on extrapolations of hadronic physics, which are poorly understood
because they lie well beyond the observations made at the Large Hadron
Collider. Magnetic fields are not well constrained by data, but if we adopt recent
models of the Galactic magnetic field~\cite{7,8}, typical values of the
deflections of particles crossing the Galaxy are a few tens of degrees for
$E/Z=10$\,EeV, depending on the direction considered~\cite{9}. Extragalactic
magnetic fields may also be relevant for cosmic rays propagating through
intergalactic space~\cite{10}. However, even if particles from individual
sources are strongly deflected, it remains possible that anisotropies in the
distribution of their arrival directions will be detectable on large angular
scales, provided the sources have a nonuniform spatial distribution or, in the
case of a single dominant source, if the cosmic-ray propagation is
diffusive~\cite{11,12,13,14}.

Searches for large-scale anisotropies are conventionally made by looking for
nonuniformities in the distribution of events in right ascension~\cite{15,16}
because, for arrays of detectors that operate with close to 100\% efficiency,
the total exposure as a function of this angle is almost constant.  The
nonuniformity of the detected cosmic-ray flux in declination (fig.~\ref{f:S1})
imprints a characteristic nonuniformity in the distribution of azimuth angles
in the local coordinate system of the array. From this distribution it becomes
possible to obtain information on the three components of a dipolar model.

\section*{Event observations, selection, and calibration}

We analyzed data recorded at the Pierre Auger Observatory between 1 January
2004 and 31 August 2016, from a total exposure of about
76,800\,km$^2$\,sr\,year.  The 1.2-m depth of the water-Cherenkov detectors
enabled us to record events at a useful rate out to large values of the zenith
angle, $\theta$. We selected events with $\theta<80^\circ$ enabling the
declination range $-90^\circ<\delta<45^\circ$ to be explored, thus covering
85\% of the sky. We adopted 4\,EeV as the threshold for selection; above that
energy, showers falling anywhere on the array are detected with 100\%
efficiency~\cite{17}. The arrival directions of cosmic rays were determined
from the relative arrival times of the shower front at each of the triggered
detectors; the angular resolution was better than $1^\circ$ at the energies
considered here~\cite{5}.

Two methods of reconstruction have been used for showers with zenith angles
above and below $60^\circ$~\cite{18,17}.  These have to account for the effects
of the geomagnetic field~\cite{19,17} and, in the case of showers with
$\theta<60^\circ$, also for atmospheric effects~\cite{20} because systematic
modulations to the rates could otherwise be induced (see supplementary
materials). The energy estimators for both data sets were calibrated
using events detected simultaneously by the water-Cherenkov detectors and the
fluorescence telescopes, with a quasi-calorimetric determination of the energy
coming from the fluorescence measurements. The statistical uncertainty in the
energy determination is 16\% above 4\,EeV and 12\% above 10\,EeV, whereas the
systematic uncertainty on the absolute energy scale, common to both data sets,
is 14\%~\cite{21}. Evidence that the analyses of the events with
$\theta<60^\circ$ and of those with $60^\circ<\theta<80^\circ$ are consistent
with each other comes from the energy spectra determined for the two angular
bands. The spectra agree within the statistical uncertainties over the energy
range of interest~\cite{22}.

We consider events in two energy ranges, $4\,\text{EeV}<E<8$\,EeV and
$E\geq8$\,EeV, as adopted in previous analyses (e.g., \cite{23,24,25}). The bin
limits follow those chosen previously in~\cite{26,27}. The median energies for
these bins are 5.0\,EeV and 11.5\,EeV, respectively. In earlier
work~\cite{23,24,25}, the event selection required that the station with the
highest signal be surrounded by six operational detectors---a demanding
condition. The number of triggered stations is greater than four for 99.2\% of
all events above 4\,EeV and for 99.9\% of events above 8\,EeV, making it
possible to use events with only five active detectors around the one with the
largest signal.  With this more relaxed condition, the effective exposure is
increased by 18.5\%, and the total number of events increases correspondingly
from 95,917 to 113,888. The reconstruction accuracy for the additional events
is sufficient for our analysis (see supplementary materials and
fig.~\ref{f:S4}).

\section*{Rayleigh analysis in right ascension}

A standard approach for studying the large-scale an\-iso\-tropies in the arrival
directions of cosmic rays is to perform a harmonic analysis in right ascension,
$\alpha$. The first-harmonic Fourier components are given by
\begin{equation}
a_\alpha =
  \frac{2}{\mathcal{N}}
  \sum _{i=1}^N w_i \cos\alpha_i,
\qquad
b_\alpha =
  \frac{2}{\mathcal{N}}
  \sum_{i=1}^N w_i \sin\alpha_i.
\label{e:1}
\end{equation}
The sums run over all $N$ detected events, each with right ascension
$\alpha_i$, with the normalization factor $\mathcal{N}=\sum_{i=1}^N w_i$. The
weights, $w_i$, are introduced to account for small nonuniformities in the
exposure of the array in right ascension and for the effects of a tilt of the
array towards the southeast (see supplementary materials). The average tilt
between vertical and the normal to the plane on which the detectors are
deployed is $0.2^\circ$, so that the effective area of the array is slightly
larger for showers arriving from the downhill direction. This introduces a
harmonic dependence in azimuth of amplitude $0.3\%\times\tan\theta$ to the
exposure. The effective aperture of the array is determined every minute.
Because the exposure has been accumulated over more than 12 years, the total
aperture is modulated by less than ${\sim}0.6\%$ as the zenith of the
observatory moves in right ascension. Events are weighted by the inverse of the
relative exposure to correct these effects (fig.~\ref{f:S2}).

The amplitude $r_\alpha$ and phase $\varphi_\alpha$ of the first harmonic of
the modulation are obtained from
\begin{equation}
r_\alpha =
  \sqrt{a_\alpha^2 + b_\alpha^2},
\qquad
\tan\varphi_\alpha =
  \frac{b_\alpha}{a_\alpha}.
\end{equation}
Table~\ref{t:1} shows the harmonic amplitudes and phases for both energy
ranges.  The statistical uncertainties in the Fourier amplitudes are
$\sqrt{2/\mathcal{N}}$; the uncertainties in the amplitude and phase correspond
to the 68\% confidence level of the marginalized probability distribution
functions.  The rightmost column shows the probabilities that amplitudes larger
than those observed could arise by chance from fluctuations in an isotropic
distribution.  These probabilities are calculated as
$P(r_\alpha)=\exp(-\mathcal{N}r_\alpha^2/4)$~\cite{28}. For the lower energy
bin ($4\,\text{EeV}<E<8$\,EeV), the result is consistent with isotropy, with a
bound on the harmonic amplitude of ${<}1.2\%$ at the 95\% confidence level. For
the events with $E\geq8$\,EeV, the amplitude of the first harmonic is
$4.7_{-0.7}^{+0.8}\%$, which has a probability of arising by chance of
$2.6{\times}10^{-8}$, equivalent to a two-sided Gaussian significance of
5.6$\sigma$. The evolution of the significance of this signal with time is
shown in fig.~\ref{f:S3}; the dipole became more significant as the exposure
increased. Allowing for a penalization factor of 2 to account for the fact that
two energy bins were explored, the significance is reduced to 5.4$\sigma$.
Further penalization for the four additional lower energy bins examined
in~\cite{23} has a similarly mild impact on the significance, which falls to
5.2$\sigma$. The maximum of the modulation is at right ascension of
$100^\circ\pm10^\circ$. The maximum of the modulation for the
$4\,\text{EeV}<E<8$\,EeV bin, at $80^\circ\pm60^\circ$, is compatible with the
one determined in the higher-energy bin, although it has high uncertainty and
the amplitude is not statistically significant. Table~\ref{t:S1} shows that
results obtained under the stricter trigger condition and for the additional
events gained after relaxing the trigger are entirely consistent with each
other.

\begin{table}[t]
\caption{\textbf{First-harmonic in right ascension.} Data are from the Rayleigh
analysis of the first harmonic in right ascension for the two energy bins.}
\label{t:1}
\begin{center}
\begin{tabular}{llrrrrr}
\toprule
\textbf{\makecell{Energy\\{}[EeV]}} &
  \textbf{\makecell{Number\\of events}} &
  \textbf{\makecell{Fourier\\coefficient $a_\alpha$}} & 
  \textbf{\makecell{Fourier\\coefficient $b_\alpha$}} &
  \textbf{\makecell{Amplitude\\$r_\alpha$}} &
  \textbf{\makecell{Phase $\varphi_\alpha$\\{}[$^\circ$]}} &
  \textbf{\makecell{Probability\\$P({\geq}r_\alpha)$}}
\\
\midrule
4 to 8 & 81,701 & $0.001\pm0.005$ & $0.005\pm0.005$ & $0.005_{-0.002}^{+0.006}$ & $80\pm60$ & 0.60
\\[2mm]
${\geq}8$ & 32,187 & $-0.008\pm0.008$ & $0.046\pm0.008$ & $0.047_{-0.007}^{+0.008}$ & $100\pm10$ & $2.6{\times}10^{-8}$
\\
\bottomrule
\end{tabular}
\end{center}
\end{table}

\begin{figure}[t]
\centering
\includegraphics[width=0.6\textwidth]{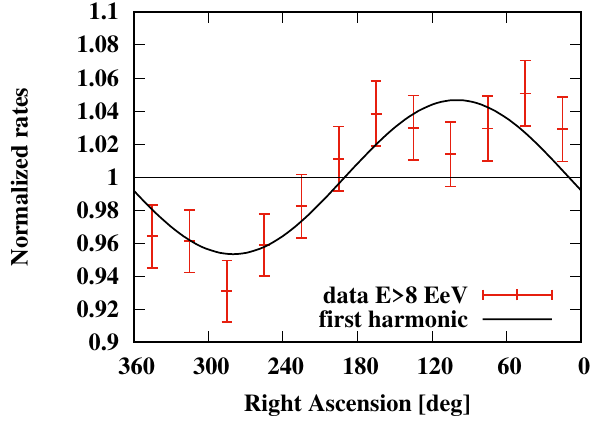}
\caption{\textbf{Normalized rate of events as a function of right ascension.}
Normalized rate for 32,187 events with $E\geq8$\,EeV, as a function of right
ascension (integrated in declination). Error bars are 1$\sigma$ uncertainties.
The solid line shows the first-harmonic modulation from Table~\ref{t:1}, which
displays good agreement with the data ($\chi^2/n=10.5/10$); the dashed line
shows a constant function.}
\label{f:1}
\end{figure}

Figure~\ref{f:1} shows the distribution of the normalized rate of events
above 8\,EeV as a function of right ascension. The sinusoidal function
corresponds to the first harmonic; the distribution is compatible
with a dipolar modulation: $\chi^2/n=10.5/10$ for the first-harmonic curve and
$\chi^2/n=45/12$ for a constant function (where $n$ is the number of degrees of
freedom, equal to the number of points in the plot minus the number of
parameters of the fit).

\begin{figure}[t]
\centering
\includegraphics[width=0.9\textwidth]{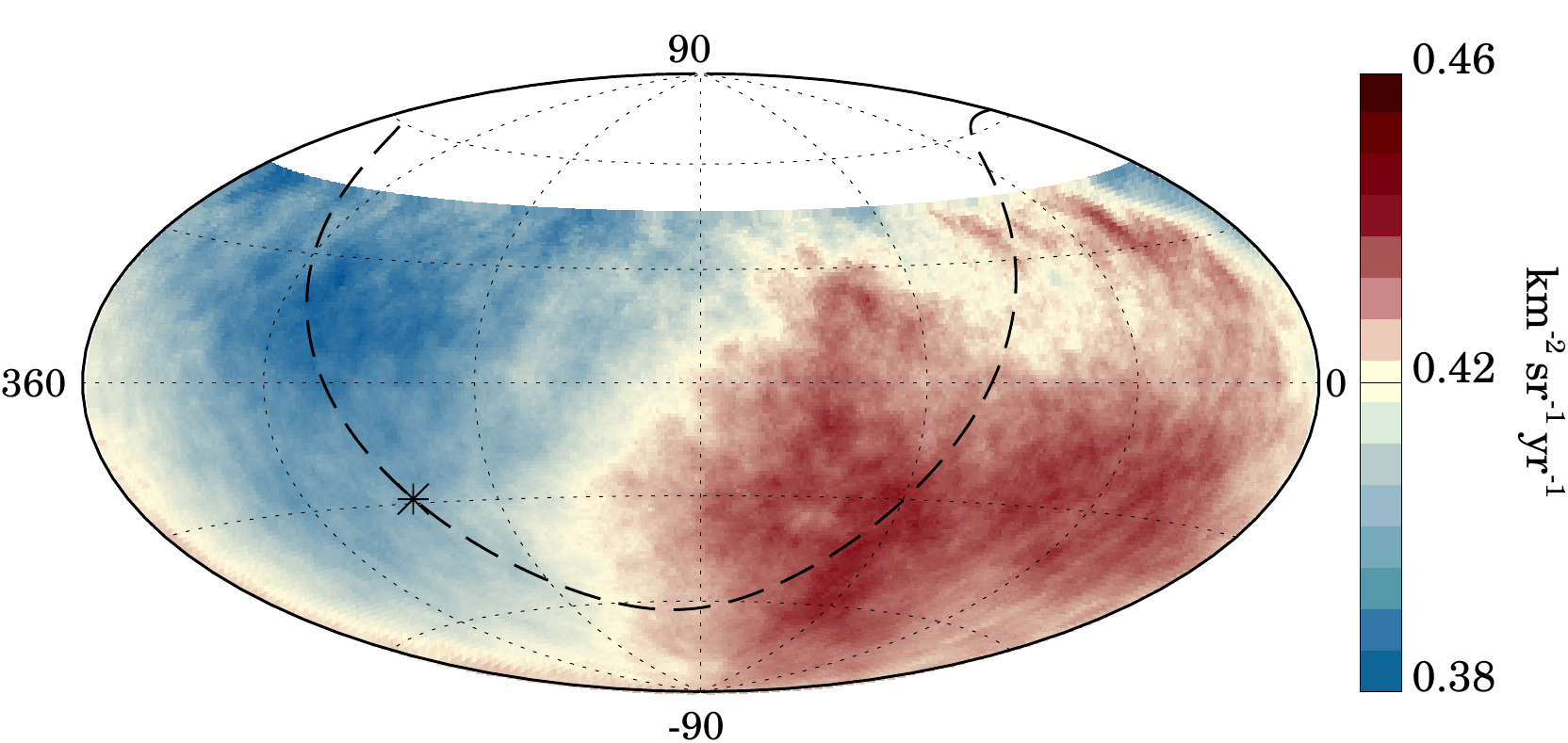}
\caption{\textbf{Map showing the fluxes of particles in equatorial
coordinates.} Sky map in equatorial coordinates, using a Hammer projection,
showing the cosmic-ray flux above 8\,EeV smoothed with a $45^\circ$ top-hat
function. The Galactic center is marked with an asterisk and the Galactic plane
is shown by a dashed line.}
\label{f:2}
\end{figure}

The distribution of events in equatorial coordinates, smoothed with a
$45^\circ$ radius top-hat function to better display the large-scale features,
is shown in Fig.~\ref{f:2}.

\section*{Reconstruction of the three-dimensional dipole}

In the presence of a three-dimensional dipole, the Rayleigh analysis in right
ascension is sensitive only to its component orthogonal to the rotation axis of
Earth, $d_\perp$. A dipole component in the direction of the rotation axis
of Earth, $d_z$, induces no modulation of the flux in right ascension, but
does so in the azimuthal distribution of the directions of arrival at the
array. A non-vanishing value of $d_z$ leads to a sinusoidal modulation in
azimuth with a maximum toward the northern or the southern direction.

To recover the three-dimensional dipole, we combine the first-harmonic analysis
in right ascension with a similar one in the azimuthal angle $\varphi$,
measured counterclockwise from the east. The relevant component, $b_\varphi$,
is given by an expression analogous to that in Eq.~\ref{e:1}, but in terms of
the azimuth of the arrival direction of the shower rather than in terms of the
right ascension. The results are $b_\varphi=-0.013\pm0.005$ in the
$4\,\text{EeV}<E<8$\,EeV bin and $b_\varphi=-0.014\pm0.008$ in the
$E\geq8$\,EeV bin. The probabilities that larger or equal absolute values for
$b_\varphi$ arise from an isotropic distribution are 0.8\% and 8\%,
respectively.

Under the assumption that the dominant cosmic-ray anisotropy is dipolar, based
on previous studies that found that the effects of higher-order multipoles are
not significant in this energy range~\cite{25,29,30}, the dipole components and
its direction in equatorial coordinates $(\alpha_\text{d}, \delta_\text{d})$
can be estimated from
\begin{equation}
d_\perp \simeq
  \frac{r_\alpha}{\avg{\cos\delta}},
\qquad
d_z \simeq
  \frac{b_\varphi}{\cos\ell_\text{obs} \avg{\sin\theta}},
\qquad
\alpha_\text{d} =
  \varphi_\alpha,
\qquad
\tan\delta_\text{d} =
  \frac{d_z}{d_\perp},
\end{equation}
\cite{25}, where $\avg{\cos\delta}$ is the mean cosine of the declinations of
the events, $\avg{\sin\theta}$ is the mean sine of the zenith angles of the
events, and $\ell_{obs}\simeq-35.2^\circ$ is the average latitude of the
Observatory. For our data set, we find $\avg{\cos\delta}=0.78$ and
$\avg{\sin\theta}=0.65$.

The parameters describing the direction of the three-dimensional dipole are
summarized in Table~\ref{t:2}. For $4\,\text{EeV}<E<8$\,EeV, the dipole
amplitude is $d=2.5_{-0.7}^{+1.0}\%$, pointing close to the celestial south
pole, at $(\alpha_\text{d},\delta_\text{d})=(80^\circ,-75^\circ)$, although the
amplitude is not statistically significant. For energies above 8\,EeV, the
total dipole amplitude is $d=6.5_{-0.9}^{+1.3}\%$, pointing toward
$(\alpha_\text{d},\delta_\text{d})=(100^\circ,-24^\circ)$.  In Galactic
coordinates, the direction of this dipole is $(\ell,b)=(233^\circ,-13^\circ)$.
This dipolar pattern is clearly seen in the flux map in Fig.~\ref{f:2}. To
establish whether the departures from a perfect dipole are just statistical
fluctuations or indicate the presence of additional structures at smaller
angular scales would require at least twice as many events.

\begin{table}[t]
\caption{\textbf{Three dimensional dipole reconstruction.} Directions of dipole
components are shown in equatorial coordinates.}
\label{t:2}
\begin{center}
\resizebox{\columnwidth}{!}{
\begin{tabular}{cccccc}
\toprule
\textbf{\makecell{Energy\\{}[EeV]}} &
  \textbf{\makecell{Dipole\\component $d_z$}} &
  \textbf{\makecell{Dipole\\component $d_\perp$}} &
  \textbf{\makecell{Dipole\\amplitude $d$}} &
  \textbf{\makecell{Dipole\\declination $\delta_\mathrm{d}$ [$^\circ$]}} &
  \textbf{\makecell{Dipole right\\ascension $\alpha_\mathrm{d}$ [$^\circ$]}}
\\
\midrule
4 to 8 & $-0.024\pm0.009$ & $0.006_{-0.003}^{+0.007}$ & $0.025_{-0.007}^{+0.010}$ & $-75_{-8}^{+17}$ & $80\pm60$
\\[2mm]
8 & $-0.026\pm0.015$ & $0.060_{-0.010}^{+0.011}$ & $0.065_{-0.009}^{+0.013}$ & $-24_{-13}^{+12}$ & $100\pm10$
\\
\bottomrule
\end{tabular}
}
\end{center}
\end{table}

\section*{Implications for the origin of high-energy cosmic rays}

The anisotropy we have found should be seen in the context of related results
at lower energies. Above a few PeV, the steepening of the cosmic ray energy
spectrum has been interpreted as being due to efficient escape of particles
from the Galaxy and/or because of the inability of the sources to accelerate
cosmic rays beyond a maximum value of $E/Z$. The origin of the particles
remains unknown. Although supernova remnants are often discussed as sources,
evidence has been reported for a source in the Galactic center capable of
accelerating particles to PeV energies~\cite{31}. Diffusive escape from the
Galaxy is expected to lead to a dipolar component with a maximum near the
Galactic center direction~\cite{32}. This is compatible with results obtained
in the $10^{15}$ to $10^{18}$\,eV range \cite{15,16,23,24,33}, which provide
values for the phase in right ascension close to that of the Galactic center,
$\alpha_\text{GC}=266^\circ$.

Models proposing a Galactic origin up to the highest observed
energies~\cite{34,35} are in increasing tension with observations. If the
Galactic sources postulated to accelerate cosmic rays above EeV energies, such
as short gamma-ray bursts or hypernovae, were distributed in the disk of the
Galaxy, a dipolar component of anisotropy is predicted with an amplitude that
exceeds existing bounds at EeV energies~\cite{24,33}. In this sense, the
constraint obtained here on the dipole amplitude (Table~\ref{t:2}) for
$4\,\text{EeV}<E<8$\,EeV further disfavors a predominantly Galactic origin.
This tension could be alleviated if cosmic rays at a few EeV were dominated by
heavy nuclei such as iron, but this would be in disagreement with the lighter
composition inferred observationally at these energies \cite{6}. The maximum of
the flux might be expected to lie close to the Galactic center region, whereas
the direction of the three-dimensional dipole determined above 8\,EeV lies
${\sim}125^\circ$ from the Galactic center. This suggests that the anisotropy
observed above 8\,EeV is better explained in terms of an extragalactic origin.
Above 40\,EeV, where the propagation should become less diffusive, there are no
indications of anisotropies associated with either the Galactic center or the
Galactic plane~\cite{36}.

There have been many efforts to interpret the properties of ultrahigh-energy
cosmic rays in terms of extragalactic sources. Because of Liouville's theorem,
the distribution of cosmic rays must be anisotropic outside of the Galaxy for
an anisotropy to be observed at Earth. An anisotropy cannot arise through
deflections of an originally isotropic flux by a magnetic field. One prediction
of anisotropy comes from the Compton-Getting effect~\cite{37}, which results
from the proper motion of the Earth in the rest frame of cosmic-ray sources,
but the amplitude is expected to be only 0.6\%~\cite{38}, well below what has
been observed. Other studies have predicted larger anisotropies. These assume
that ultrahigh-energy cosmic rays originate from an inhomogeneous distribution
of sources~\cite{13,14,39}, or that they arise from a dominant source and then
diffuse through intergalactic magnetic fields~\cite{11,12,13,14}.  The
resulting dipole amplitudes are predicted to grow with energy, reaching 5 to
20\% at 10\,EeV.  These amplitudes depend on the cosmic-ray composition as well
as the details of the source distribution. On average, the predictions are
smaller for larger source densities or for more isotropically distributed
sources. If the sources were distributed like galaxies, the distribution of
which has a significant dipolar component~\cite{40}, a dipolar cosmic-ray
anisotropy would be expected in a direction similar to that of the dipole
associated with the galaxies. This effect would be due to the excess of
cosmic-ray sources in this direction and is different from the Compton-Getting
effect due to the Earth's motion with respect to the rest frame of cosmic rays.
For the infrared-detected galaxies in the 2MRS catalogue~\cite{40}, the
flux-weighted dipole points in Galactic coordinates in the direction
$(\ell,b=(251^\circ,38^\circ)$. In this coordinate system, the dipole we detect
for cosmic rays above 8\,EeV is in the direction $(233^\circ,-13^\circ)$, about
$55^\circ$ away from that of the 2MRS dipole.

\begin{figure}[t]
\centering
\includegraphics[width=0.9\textwidth]{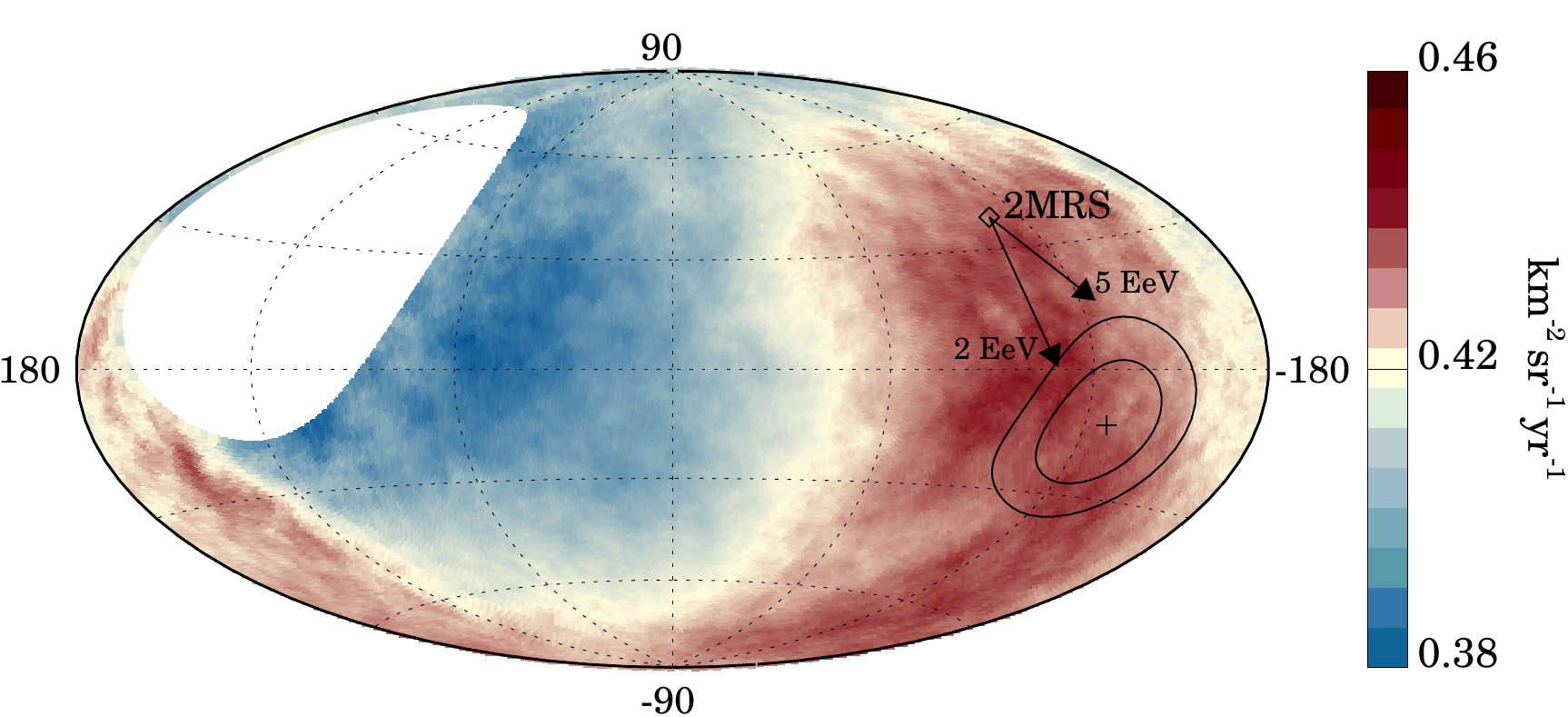}
\caption{\textbf{Map showing the fluxes of particles in Galactic coordinates.}
Sky map in Galactic coordinates showing the cosmic-ray flux for $E\geq8$\,EeV
smoothed with a $45^\circ$ top-hat function. The Galactic center is at the
origin. The cross indicates the measured dipole direction; the contours denote
the 68\% and 95\% confidence-level regions. The dipole in the 2MRS galaxy
distribution is indicated. Arrows show the deflections expected for a
particular model of the Galactic magnetic field~\cite{8} on particles with
$E/Z=5$\,EeV or 2\,EeV.}
\label{f:3}
\end{figure}

For an extragalactic origin, the Galactic magnetic fields modify the direction
of the dipole observed at Earth relative to its direction outside the Galaxy. For illustration, Fig.~\ref{f:3} shows a map of the flux above 8\,EeV
in which the direction of the cosmic-ray dipole is shown along with the
direction towards the 2MRS dipole. The arrows in the plot indicate how a
dipolar distribution of cosmic rays, in the same direction as the 2MRS dipole
outside the Galaxy, would be affected by the Galactic magnetic field~\cite{8}.
The tips of the arrows indicate the direction of the dipole of the flux
arriving at Earth, assuming common values of $E/Z=5$\,EeV or 2\,EeV. Given
the inferred average values for $Z\sim1.7$ to 5 at 10\,EeV, these represent
typical values of $E/Z$ for the cosmic rays contributing to the observed
dipole. The agreement between the directions of the dipoles is improved by
adopting these assumptions about the charge composition and the deflections in
the Galactic magnetic field. For these directions, the deflections within the
Galaxy will also lead to a lowering of the amplitude of the dipole to about
90\% and 70\% of the original value, for $E/Z=5$\,EeV and 2\,EeV, respectively.
The lower amplitude in the $4\,\text{EeV}<E<8$\,EeV bin might also be the
result of stronger magnetic deflections at lower energies.

Our findings constitute the observation of an anisotropy in the arrival
direction of cosmic rays with energies above 8\,EeV. The anisotropy can be well
represented by a dipole with an amplitude of $6.5_{-0.9}^{+1.3}\%$ in the
direction of right ascension $\alpha_\text{d}=100\pm10^\circ$ and declination
$\delta_\text{d}={-24_{-13}^{+12}}^\circ$. By comparing our results with
phenomenological predictions, we find that the magnitude and direction of the
anisotropy support the hypothesis of an extragalactic origin for the
highest-energy cosmic rays, rather than sources within the Galaxy.

\renewcommand{\refname}{References and Notes:}

\section*{Acknowledgments}

The successful installation, commissioning, and operation of the Pierre Auger
Observatory would not have been possible without the strong commitment from the
technical and administrative staff in Malarg\"ue, and the financial support
from a number of funding agencies in the participating countries. The full
acknowledgments are in the Supplementary Materials. The Pierre Auger
Collaboration will make public the data reproduced in the Figures of this paper
on the Auger website on the link to this publication:
\href{http://www.auger.org/data/science2017.tar.gz}{www.auger.org/data/science2017.tar.gz}

\clearpage

\section*{Full author list and affiliations}

$\null$
\begin{wrapfigure}[10]{l}{0.12\linewidth}
\includegraphics[width=0.98\linewidth,clip=]{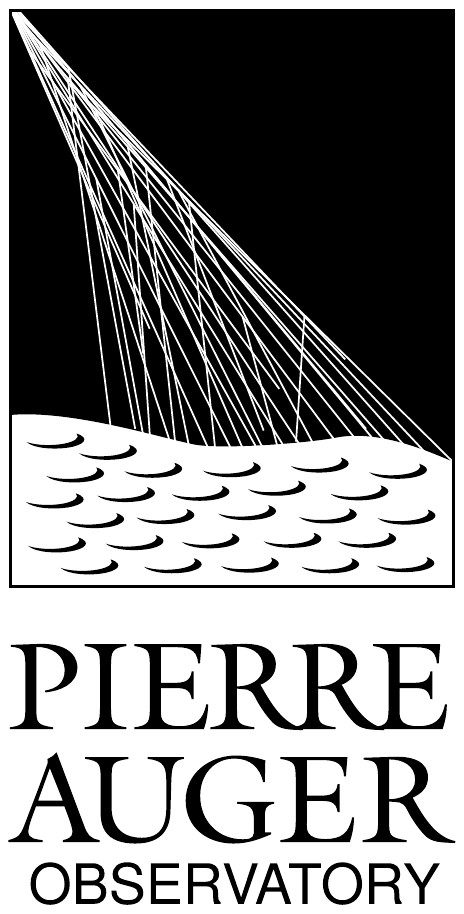}
\end{wrapfigure}
\begin{sloppypar}\noindent
\input{latex_authorlist_authors}
\end{sloppypar}

\vspace{1ex}
\begin{center}
\rule{0.1\columnwidth}{0.5pt}
\raisebox{-0.4ex}{\scriptsize$\bullet$}
\rule{0.1\columnwidth}{0.5pt}
\end{center}

\vspace{1ex}
\input{latex_authorlist_institutions}

\section*{Acknowledgments}

The successful installation, commissioning, and operation of the Pierre Auger
Observatory would not have been possible without the strong commitment and
effort from the technical and administrative staff in Malarg\"ue.

\begin{sloppypar}
We are very grateful to the following agencies and organizations for financial
support: Argentina -- Comisi\'on Nacional de Energ\'ia At\'omica; Agencia
Nacional de Promoci\'on Cient\'ifica y Tecnol\'ogica (ANPCyT); Consejo Nacional
de Investigaciones Cient\'ificas y T\'ecnicas (CONICET); Gobierno de la
Provincia de Mendoza; Municipalidad de Malarg\"ue; NDM Holdings and Valle Las
Le\~nas; in gratitude for their continuing cooperation over land access;
Australia -- the Australian Research Council; Brazil -- Conselho Nacional de
Desenvolvimento Cient\'ifico e Tecnol\'ogico (CNPq); Financiadora de Estudos e
Projetos (FINEP); Funda\c{c}\~ao de Amparo \`a Pesquisa do Estado de Rio de
Janeiro (FAPERJ); S\~ao Paulo Research Foundation (FAPESP) Grants
No.\ 2010/07359-6 and No.~1999/05404-3; Minist\'erio de Ci\^encia e Tecnologia
(MCT); Czech Republic -- Grant No.~MSMT CR LG15014, LO1305, LM2015038 and
CZ.02.1.01/0.0/0.0/16\_013/0001402; France -- Centre de Calcul IN2P3/CNRS;
Centre National de la Recherche Scientifique (CNRS); Conseil R\'egional
Ile-de-France; D\'epartement Physique Nucl\'eaire et Corpusculaire
(PNC-IN2P3/CNRS); D\'epartement Sciences de l'Univers (SDU-INSU/CNRS); Institut
Lagrange de Paris (ILP) Grant No.~LABEX ANR-10-LABX-63 within the
Investissements d'Avenir Programme Grant No.~ANR-11-IDEX-0004-02; Germany --
Bundesministerium f\"ur Bildung und Forschung (BMBF); Deutsche
Forschungsgemeinschaft (DFG); Finanzministerium Baden-W\"urttemberg; Helmholtz
Alliance for Astroparticle Physics (HAP); Helmholtz-Gemeinschaft Deutscher
Forschungszentren (HGF); Ministerium f\"ur Innovation, Wissenschaft und
Forschung des Landes Nordrhein-Westfalen; Ministerium f\"ur Wissenschaft,
Forschung und Kunst des Landes Baden-W\"urttemberg; Italy -- Istituto Nazionale
di Fisica Nucleare (INFN); Istituto Nazionale di Astrofisica (INAF); Ministero
dell'Istruzione, dell'Universit\'a e della Ricerca (MIUR); CETEMPS Center of
Excellence; Ministero degli Affari Esteri (MAE); Mexico -- Consejo Nacional de
Ciencia y Tecnolog\'ia (CONACYT) No.~167733; Universidad Nacional Aut\'onoma de
M\'exico (UNAM); PAPIIT DGAPA-UNAM; The Netherlands -- Ministerie van
Onderwijs, Cultuur en Wetenschap; Nederlandse Organisatie voor Wetenschappelijk
Onderzoek (NWO); Stichting voor Fundamenteel Onderzoek der Materie (FOM);
Poland -- National Centre for Research and Development, Grants
No.~ERA-NET-ASPERA/01/11 and No.~ERA-NET-ASPERA/02/11; National Science Centre,
Grants No.~2013/08/M/ST9/00322, No.~2013/08/M/ST9/00728 and No.~HARMONIA
5--2013/10/M/ST9/00062, UMO-2016/22/M/ST9/00198; Portugal -- Portuguese
national funds and FEDER funds within Programa Operacional Factores de
Competitividade through Funda\c{c}\~ao para a Ci\^encia e a Tecnologia
(COMPETE); Romania -- Romanian Authority for Scientific Research ANCS;
CNDI-UEFISCDI partnership projects Grants No.~20/2012 and No.194/2012 and PN 16
42 01 02; Slovenia -- Slovenian Research Agency; Spain -- Comunidad de Madrid;
Fondo Europeo de Desarrollo Regional (FEDER) funds; Ministerio de Econom\'ia y
Competitividad; Xunta de Galicia; European Community 7th Framework Program
Grant No.~FP7-PEOPLE-2012-IEF-328826; USA -- Department of Energy, Contracts
No.~DE-AC02-07CH11359, No.~DE-FR02-04ER41300, No.~DE-FG02-99ER41107 and
No.~DE-SC0011689; National Science Foundation, Grant No.~0450696; The Grainger
Foundation; Marie Curie-IRSES/EPLANET; European Particle Physics Latin American
Network; European Union 7th Framework Program, Grant No.~PIRSES-2009-GA-246806;
European Union's Horizon 2020 research and innovation programme (Grant
No.~646623); and UNESCO.
\end{sloppypar}

\clearpage

\appendix

\renewcommand{\theequation}{S\arabic{equation}}
\setcounter{equation}{0}
\renewcommand\thefigure{S\arabic{figure}}
\setcounter{figure}{0}
\renewcommand\thetable{S\arabic{table}}
\setcounter{table}{0}

\clearpage

\section*{Supplementary Material}

\section*{Materials and Methods}

\subsection*{Event weights}

\begin{figure}[t]
\centering
\includegraphics[width=0.6\textwidth]{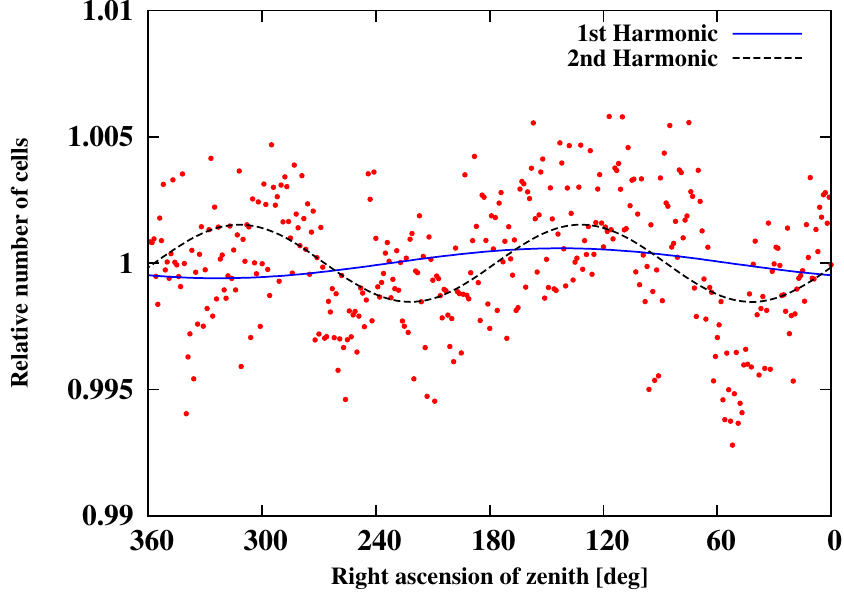}
\caption{The normalized number of active cells as a function of the right
ascension of the zenith of the observatory.  Data are shown for the time period
2004 January 1 to 2016 August 31. The best-fitting first and second harmonics
are overlaid. The first harmonic has an amplitude of $(0.06\pm0.02)\%$ and the
second harmonic has an amplitude of $(0.15\pm0.02)\%$.}
\label{f:S2}
\end{figure}

The Fourier harmonic analysis we perform accounts for modulations in the
exposure (of instrumental origin) as well as the effects due to the tilt of the
surface detector array. To achieve this each event is weighted by a factor
\begin{equation}
w_i =
  \left[
    \left(
      1 +
      \sin\theta_\text{tilt} \,
      \tan\theta_i \,
      \cos\left(
        \varphi_i - \varphi_\text{tilt}
      \right)
    \right)
    \Delta N_\text{cell}(\alpha_i^0)
  \right]^{-1},
\label{e:S1}
\end{equation}
where $\theta_i$ and $\phi_i$ are the zenith and azimuth angles of the events
and $\alpha_i^0$ is the right ascension directly above the array at the time
the $i$-th event was detected. The term enclosed by the round brackets accounts
for the effects of the slight tilt of the array. The average inclination from
the vertical is $\theta_\text{tilt}\approx0.2^\circ$, in the direction
$\phi_\text{tilt}\approx-30^\circ$, i.e.\ $30^\circ$ South of the Easterly
direction. The tilt term affects only the Fourier analysis in azimuth, and thus
the dipole component $d_z$. The second term in Eq.~\eqref{e:S1}, $\Delta
N_\text{cell}(\alpha_0)$, allows for the fact that the effective aperture of
the observatory is not uniform in sidereal time. This factor corresponds to the
relative number of detector cells, i.e.\ the active detectors surrounded by at
least five other active detectors, present when the right ascension of the
zenith of the observatory equals $\alpha_0$ within binning accuracy. It is
obtained by adding the number of cells over the whole period of observations,
with dead times due to power failures or to communication or acquisition
problems discarded. The total number of cells within each $\alpha_0$ bin is
normalized to the average value~\cite{22}.  $\Delta N_\text{cell}(\alpha_0)$ is
plotted in Fig.~\ref{f:S2} with a bin width of $1^\circ$.  This term affects
only the Fourier analysis in right ascension, and thus the dipole component
$d_\perp$. After more than 12 years of continuous operation of the observatory
the normalized number of cells shows variations that are smaller than
$\pm0.6\%$. If the effects of the modulation in the number of cells were not
taken into account through the weights, a spurious contribution to $d_\perp$ of
amplitude 0.05\% in the direction $\alpha\approx145^\circ$ would be induced.
This contribution is an order of magnitude smaller than the statistical
uncertainty in the determination of this dipole component, having then a
marginal effect. If the effects of the tilt were not taken into account, a
spurious contribution to $d_z$ of $-0.4\%$ would be induced. Had we not
introduced weighting to the Fourier analysis we would have obtained: for
$4\,\text{EeV}<E<8$\,EeV, a total dipole amplitude $d=2.9_{-0.8}^{+1.0}\%$ at
$(\alpha_\text{d},\delta_\text{d})=(85^\circ,-77^\circ)$ with corresponding
values for $E\geq8$\,EeV of $d=6.7_{-0.9}^{+1.3}\%$ at
$\alpha_\text{d},\delta_\text{d})=(100^\circ,-26^\circ)$.

\subsection*{Energy reconstruction}

Two methods of reconstruction have been used for showers with zenith angles
above and below $60^\circ$~\cite{16,17}. The energy estimator used for showers
with $\theta<60^\circ$ is the signal reconstructed at 1000\,m from the shower
core. This signal is corrected for atmospheric effects~\cite{18} that would
otherwise introduce systematic modulations to the rates as a function of time
of day or season. This could result in spurious influences on the distribution
in sidereal time (a time scale that is based on the Earth's rate of rotation
measured relative to the fixed stars rather than the Sun, corresponding to
366.25\,cycles/year) and hence could be a source of systematic effects for the
anisotropies inferred. The atmospheric effects arise from the dependences of
the longitudinal and lateral attenuation of the electromagnetic component of
air showers on atmospheric conditions, in particular temperature and pressure.
If not corrected, these could cause a modulation of the rates of up to
$\pm1.7\%$ in solar time. The energy estimator is also corrected for
geomagnetic effects~\cite{19} as otherwise a systematic modulation of amplitude
${\sim}0.7\%$ would be induced in the azimuthal distributions.

The particles arriving at the ground in showers with $\theta>60^\circ$ are
predominantly muons.  As the atmospheric thickness traversed by a shower is
proportional to $\sec\theta$, at those zenith angles the electromagnetic
component is almost completely absorbed so that atmospheric effects are
negligible. For these large angles the energy estimator is based on the muon
content relative to that in a simulated proton-shower of 10\,EeV, with the
geomagnetic deflections of muons accounted for in the reconstruction of these
events~\cite{17}.

After applying the corrections to those events with $\theta<60^\circ$, the
amplitude of modulation in solar time (365.25\,cycles/year) for the whole data
set (with $\theta<80^\circ$ and $E>4$\,EeV) is reduced to $0.5\pm0.4\%$. This
is obtained from the first harmonic Fourier analysis of the arrival times as a
function of the hour of the day. The residual effect in right ascension, after
averaging over more than 12 years, is less than one part in a thousand. As a
further check, the amplitude of modulation at the anti-sidereal frequency
(364.25\,cycles/year) is $0.5\pm0.4\%$, consistent with the absence of residual
systematic effects. The results of the solar and anti-sidereal amplitudes in
the two separate energy bins are also consistent with the absence of systematic
effects, being, for $4\,\text{EeV}<E<8$\,EeV of $0.6\pm0.5\%$ in solar and of
$0.4\pm0.5\%$ in anti-sidereal, while for $E>8$\,EeV they are $0.7\pm0.8\%$ in
solar and of $1.1\pm0.8\%$ in anti-sidereal.

\subsection*{Accuracy of the reconstruction of events obtained with
relaxed trigger}

\begin{figure}[t]
\centering
\includegraphics[width=0.6\textwidth]{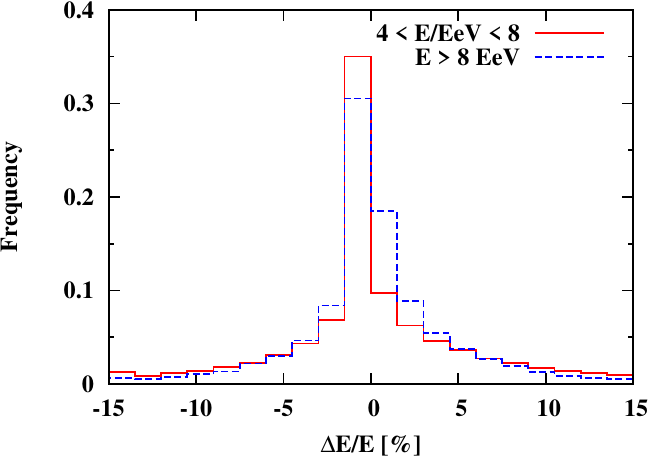}
\\
\includegraphics[width=0.6\textwidth]{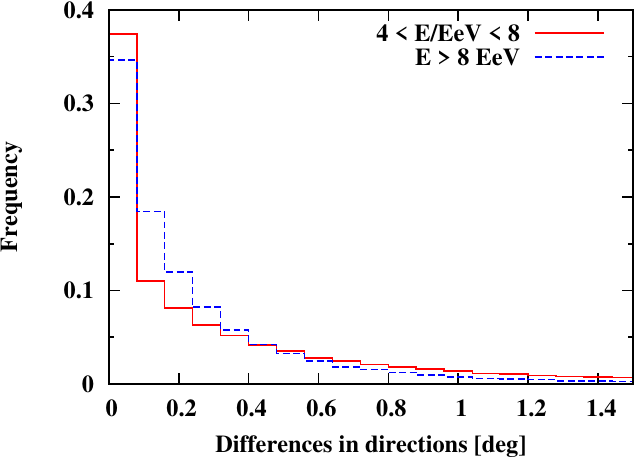} 
\caption{The difference between measurements of energy and direction for the
strict and relaxed trigger conditions. Top panel: Distribution of relative difference between the reconstructed
energies of the event satisfying the strict trigger condition and those obtained by artificially removing one of the
six detectors surrounding the one with the largest signal. Bottom panel: corresponding distribution of the angular
separation between the reconstructed arrival directions. For $E\geq8$\,EeV ($4\,\text{EeV}<E<8$\,EeV)
the difference between the reconstructed directions has an average of $0.3^\circ$ ($0.4^\circ$), with 90\% of the
events having an angular difference smaller than $0.7^\circ$ ($1.2^\circ$).
The energy estimates differ on average by only 0.2\% (0.3\%), with a dispersion of ${\sim}5\%$ (8\%).}
\label{f:S4}
\label{f:S5}
\end{figure}

Taking events passing the stricter cuts (with six active detectors surrounding
the one with the highest signal) and re-analyzing them after removing one of
the six detectors, we find that for $E\geq8$\,EeV ($4\,\text{EeV}<E<8$\,EeV)
the difference between the reconstructed directions has an average of
$0.3^\circ$ ($0.4^\circ$), with 90\% of the events having an angular difference
smaller than $0.7^\circ$ ($1.2^\circ$).  The energy estimates differ on average
by only 0.2\% \ (0.3\%), with a dispersion of ${\sim}5\%$ (8\%). These
differences are well below the experimental uncertainties of these two
parameters. The distribution of the differences in arrival directions and
reconstructed energies between the original event satisfying the strict trigger
and those with a missing detector around the one with the highest signal are
shown in fig.~\ref{f:S4}. Events passing the stricter cut from two years of
data (2013 and 2014) were analyzed, leading to a total of artificial events
passing the relaxed trigger of 65,000 for $4\,\text{EeV}<E<8$\,EeV and 27,000
for $E\geq8$,EeV.

\begin{figure}[t]
\centering
\includegraphics[width=0.6\textwidth]{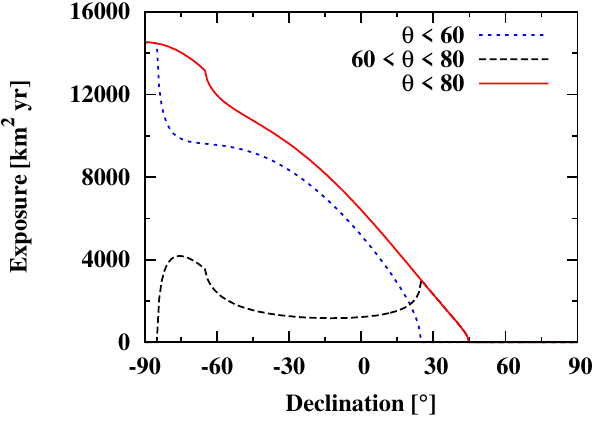}
\caption{Exposure as a function of declination. The separate contributions from
events with $\theta<60^\circ$ and $60^\circ<\theta<80^\circ$ are also displayed.}
\label{f:S1}
\end{figure}

\begin{figure}[t]
\centering
\includegraphics[width=0.6\textwidth]{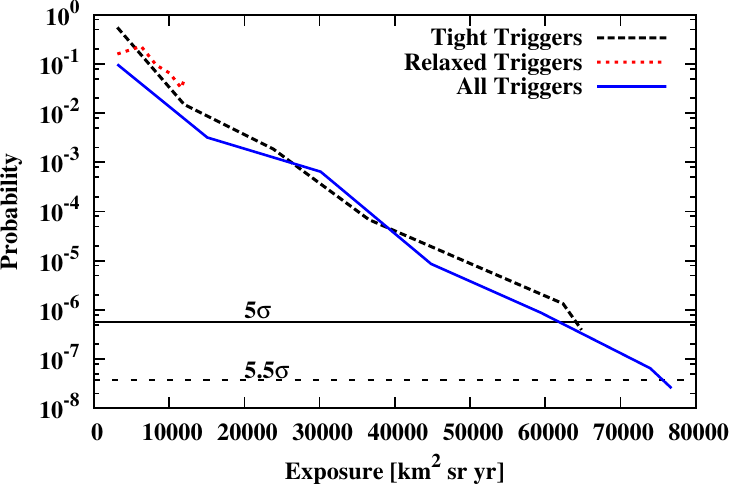}
\caption{Evolution of the probability that the signal arises by chance as a
function of time. The evolution of the probability that in an isotropic
distribution the first-harmonic amplitude in right ascension be larger or equal
than the one measured is shown for events with $E\geq8$\,EeV. This is plotted
as a function of the exposure accumulated over the years. The solid blue line
corresponds to the signal from the combination of the two triggers while the
black and red lines refer to data from the tight and relaxed triggers
respectively (Table~\ref{t:S1}). The values corresponding to 5$\sigma $ and
5.5$\sigma$ are indicated as horizontal lines.}
\label{f:S3}
\end{figure}

\begin{table}[t]
\caption{Rayleigh analysis for the first-harmonic in right ascension for
different trigger conditions in the two energy bins.}
\label{t:S1}
\begin{center}
\resizebox{\columnwidth}{!}{
\begin{tabular}{rrrrrccc}
\toprule
\textbf{Dataset} &
  \textbf{\makecell{Energy\\{}[EeV]}} &
  \textbf{\makecell{Number\\of events}} &
  \textbf{\makecell{Fourier\\coefficient $a_\alpha$}} &
  \textbf{\makecell{Fourier\\coefficient $b_\alpha$}} &
  \textbf{\makecell{Amplitude\\$r_\alpha$}} &
  \textbf{\makecell{Phase$\varphi_\alpha$\\{}[$^\circ$]}} &
  \textbf{\makecell{Probability\\$P({\geq}r_\alpha)$}}
\\
\midrule
\multirow{2}{*}{\makecell{Tight\\Triggers}} & 4 to 8 & 68,775 & $0.005\pm0.005$ & $0.005\pm0.005$ & 0.007 & $50\pm50$ & 0.44
\\
 & ${\geq}8$ & 27,142 & $-0.006\pm0.009$ & $0.046\pm0.009$ & 0.047 & $97\pm11$ & $3.9{\times}10^{-7}$
\\
\midrule
\multirow{2}{*}{\makecell{Relaxed\\Triggers}} & 4 to 8 &  12,926 & $-0.019\pm0.012$ & $0.003\pm0.012$ & 0.019 & $170\pm40$ & 0.28
\\
 & ${\geq}8$ & 5,045 & $-0.023\pm0.020$ & $0.044\pm0.020$ & 0.049 & $117\pm24$ & 0.047
\\
\midrule
\multirow{2}{*}{\makecell{All\\Triggers}} & 4 to 8 & 81,701 & $0.001\pm0.005$ & $0.005\pm0.005$ & 0.005 & $80\pm60$ & 0.60
\\
 & ${\geq}8$ & 32,187 & $-0.008\pm0.008$ & $0.046\pm0.008$ & 0.047 & $100\pm10$ & $2.6{\times}10^{-8}$
\\
\bottomrule
\end{tabular}
}
\end{center}
\end{table}

\end{document}

%% file: latex_authorlist_authors.tex
A.~Aab$^{63}$,
P.~Abreu$^{70}$,
M.~Aglietta$^{48,47}$,
I.~Al Samarai$^{30}$,
I.F.M.~Albuquerque$^{17}$,
I.~Allekotte$^{1}$,
A.~Almela$^{8,11}$,
J.~Alvarez Castillo$^{62}$,
J.~Alvarez-Mu\~niz$^{78}$,
G.A.~Anastasi$^{39,41}$
% ...

%% file: latex_authorlist_institutions.tex

\begin{description}[labelsep=0.2em,align=right,labelwidth=0.7em,labelindent=0em,leftmargin=2em,noitemsep]
\item[$^{1}$] Centro At\'omico Bariloche and Instituto Balseiro (CNEA-UNCuyo-CONICET), San Carlos de Bariloche, Argentina
\item[$^{2}$] Centro de Investigaciones en L\'aseres y Aplicaciones, CITEDEF and CONICET, Villa Martelli, Argentina
\item[$^{3}$] Departamento de F\'\i{}sica and Departamento de Ciencias de la Atm\'osfera y los Oc\'eanos, FCEyN, Universidad de Buenos Aires and CONICET, Buenos Aires, Argentina
\item[$^{4}$] IFLP, Universidad Nacional de La Plata and CONICET, La Plata, Argentina
\item[$^{5}$] Instituto de Astronom\'\i{}a y F\'\i{}sica del Espacio (IAFE, CONICET-UBA), Buenos Aires, Argentina
\item[$^{6}$] Instituto de F\'\i{}sica de Rosario (IFIR) -- CONICET/U.N.R.\ and Facultad de Ciencias Bioqu\'\i{}micas y Farmac\'euticas U.N.R., Rosario, Argentina
\item[$^{7}$] Instituto de Tecnolog\'\i{}as en Detecci\'on y Astropart\'\i{}culas (CNEA, CONICET, UNSAM), and Universidad Tecnol\'ogica Nacional -- Facultad Regional Mendoza (CONICET/CNEA), Mendoza, Argentina
\item[$^{8}$] Instituto de Tecnolog\'\i{}as en Detecci\'on y Astropart\'\i{}culas (CNEA, CONICET, UNSAM), Buenos Aires, Argentina
\item[$^{9}$] Observatorio Pierre Auger, Malarg\"ue, Argentina
\item[$^{10}$] Observatorio Pierre Auger and Comisi\'on Nacional de Energ\'\i{}a At\'omica, Malarg\"ue, Argentina
\item[$^{11}$] Universidad Tecnol\'ogica Nacional -- Facultad Regional Buenos Aires, Buenos Aires, Argentina
\item[$^{12}$] University of Adelaide, Adelaide, S.A., Australia
\item[$^{13}$] Universit\'e Libre de Bruxelles (ULB), Brussels, Belgium
\item[$^{14}$] Centro Brasileiro de Pesquisas Fisicas, Rio de Janeiro, RJ, Brazil
\item[$^{15}$] Universidade de S\~ao Paulo, Escola de Engenharia de Lorena, Lorena, SP, Brazil
\item[$^{16}$] Universidade de S\~ao Paulo, Instituto de F\'\i{}sica de S\~ao Carlos, S\~ao Carlos, SP, Brazil
\item[$^{17}$] Universidade de S\~ao Paulo, Instituto de F\'\i{}sica, S\~ao Paulo, SP, Brazil
\item[$^{18}$] Universidade Estadual de Campinas, IFGW, Campinas, SP, Brazil
\item[$^{19}$] Universidade Estadual de Feira de Santana, Feira de Santana, Brazil
\item[$^{20}$] Universidade Federal de Pelotas, Pelotas, RS, Brazil
\item[$^{21}$] Universidade Federal do ABC, Santo Andr\'e, SP, Brazil
\item[$^{22}$] Universidade Federal do Paran\'a, Setor Palotina, Palotina, Brazil
\item[$^{23}$] Universidade Federal do Rio de Janeiro, Instituto de F\'\i{}sica, Rio de Janeiro, RJ, Brazil
\item[$^{24}$] Universidade Federal Fluminense, EEIMVR, Volta Redonda, RJ, Brazil
\item[$^{25}$] Universidad Industrial de Santander, Bucaramanga, Colombia
\item[$^{26}$] Charles University, Faculty of Mathematics and Physics, Institute of Particle and Nuclear Physics, Prague, Czech Republic
\item[$^{27}$] Institute of Physics of the Czech Academy of Sciences, Prague, Czech Republic
\item[$^{28}$] Palacky University, RCPTM, Olomouc, Czech Republic
\item[$^{29}$] Institut de Physique Nucl\'eaire d'Orsay (IPNO), Universit\'e Paris-Sud, Univ.\ Paris/Saclay, CNRS-IN2P3, Orsay, France
\item[$^{30}$] Laboratoire de Physique Nucl\'eaire et de Hautes Energies (LPNHE), Universit\'es Paris 6 et Paris 7, CNRS-IN2P3, Paris, France
\item[$^{31}$] Laboratoire de Physique Subatomique et de Cosmologie (LPSC), Universit\'e Grenoble-Alpes, CNRS/IN2P3, Grenoble, France
\item[$^{32}$] Bergische Universit\"at Wuppertal, Department of Physics, Wuppertal, Germany
\item[$^{33}$] Karlsruhe Institute of Technology, Institut f\"ur Experimentelle Kernphysik (IEKP), Karlsruhe, Germany
\item[$^{34}$] Karlsruhe Institute of Technology, Institut f\"ur Kernphysik, Karlsruhe, Germany
\item[$^{35}$] Karlsruhe Institute of Technology, Institut f\"ur Prozessdatenverarbeitung und Elektronik, Karlsruhe, Germany
\item[$^{36}$] RWTH Aachen University, III.\ Physikalisches Institut A, Aachen, Germany
\item[$^{37}$] Universit\"at Hamburg, II.\ Institut f\"ur Theoretische Physik, Hamburg, Germany
\item[$^{38}$] Universit\"at Siegen, Fachbereich 7 Physik -- Experimentelle Teilchenphysik, Siegen, Germany
\item[$^{39}$] Gran Sasso Science Institute (INFN), L'Aquila, Italy
\item[$^{40}$] INAF -- Istituto di Astrofisica Spaziale e Fisica Cosmica di Palermo, Palermo, Italy
\item[$^{41}$] INFN Laboratori Nazionali del Gran Sasso, Assergi (L'Aquila), Italy
\item[$^{42}$] INFN, Sezione di Catania, Catania, Italy
\item[$^{43}$] INFN, Sezione di Lecce, Lecce, Italy
\item[$^{44}$] INFN, Sezione di Milano, Milano, Italy
\item[$^{45}$] INFN, Sezione di Napoli, Napoli, Italy
\item[$^{46}$] INFN, Sezione di Roma "Tor Vergata", Roma, Italy
\item[$^{47}$] INFN, Sezione di Torino, Torino, Italy
\item[$^{48}$] Osservatorio Astrofisico di Torino (INAF), Torino, Italy
\item[$^{49}$] Universit\`a del Salento, Dipartimento di Ingegneria, Lecce, Italy
\item[$^{50}$] Universit\`a del Salento, Dipartimento di Matematica e Fisica ``E.\ De Giorgi'', Lecce, Italy
\item[$^{51}$] Universit\`a dell'Aquila, Dipartimento di Scienze Fisiche e Chimiche, L'Aquila, Italy
\item[$^{52}$] Universit\`a di Catania, Dipartimento di Fisica e Astronomia, Catania, Italy
\item[$^{53}$] Universit\`a di Milano, Dipartimento di Fisica, Milano, Italy
\item[$^{54}$] Universit\`a di Napoli "Federico II", Dipartimento di Fisica ``Ettore Pancini``, Napoli, Italy
\item[$^{55}$] Universit\`a di Roma ``Tor Vergata'', Dipartimento di Fisica, Roma, Italy
\item[$^{56}$] Universit\`a Torino, Dipartimento di Fisica, Torino, Italy
\item[$^{57}$] Benem\'erita Universidad Aut\'onoma de Puebla, Puebla, M\'exico
\item[$^{58}$] Centro de Investigaci\'on y de Estudios Avanzados del IPN (CINVESTAV), M\'exico, D.F., M\'exico
\item[$^{59}$] Unidad Profesional Interdisciplinaria en Ingenier\'\i{}a y Tecnolog\'\i{}as Avanzadas del Instituto Polit\'ecnico Nacional (UPIITA-IPN), M\'exico, D.F., M\'exico
\item[$^{60}$] Universidad Aut\'onoma de Chiapas, Tuxtla Guti\'errez, Chiapas, M\'exico
\item[$^{61}$] Universidad Michoacana de San Nicol\'as de Hidalgo, Morelia, Michoac\'an, M\'exico
\item[$^{62}$] Universidad Nacional Aut\'onoma de M\'exico, M\'exico, D.F., M\'exico
\item[$^{63}$] IMAPP, Radboud University Nijmegen, Nijmegen, The Netherlands
\item[$^{64}$] KVI -- Center for Advanced Radiation Technology, University of Groningen, Groningen, The Netherlands
\item[$^{65}$] Nationaal Instituut voor Kernfysica en Hoge Energie Fysica (NIKHEF), Science Park, Amsterdam, The Netherlands
\item[$^{66}$] Stichting Astronomisch Onderzoek in Nederland (ASTRON), Dwingeloo, The Netherlands
\item[$^{67}$] Institute of Nuclear Physics PAN, Krakow, Poland
\item[$^{68}$] University of \L{}\'od\'z, Faculty of Astrophysics, \L{}\'od\'z, Poland
\item[$^{69}$] University of \L{}\'od\'z, Faculty of High-Energy Astrophysics,\L{}\'od\'z, Poland
\item[$^{70}$] Laborat\'orio de Instrumenta\c{c}\~ao e F\'\i{}sica Experimental de Part\'\i{}culas -- LIP and Instituto Superior T\'ecnico -- IST, Universidade de Lisboa -- UL, Lisboa, Portugal
\item[$^{71}$] ``Horia Hulubei'' National Institute for Physics and Nuclear Engineering, Bucharest-Magurele, Romania
\item[$^{72}$] Institute of Space Science, Bucharest-Magurele, Romania
\item[$^{73}$] University Politehnica of Bucharest, Bucharest, Romania
\item[$^{74}$] Experimental Particle Physics Department, J.\ Stefan Institute, Ljubljana, Slovenia
\item[$^{75}$] Laboratory for Astroparticle Physics, University of Nova Gorica, Nova Gorica, Slovenia
\item[$^{76}$] Universidad Complutense de Madrid, Madrid, Spain
\item[$^{77}$] Universidad de Granada and C.A.F.P.E., Granada, Spain
\item[$^{78}$] Universidad de Santiago de Compostela, Santiago de Compostela, Spain
\item[$^{79}$] Case Western Reserve University, Cleveland, OH, USA
\item[$^{80}$] Colorado School of Mines, Golden, CO, USA
\item[$^{81}$] Department of Physics and Astronomy, Lehman College, City University of New York, Bronx, NY, USA
\item[$^{82}$] Louisiana State University, Baton Rouge, LA, USA
\item[$^{83}$] Michigan Technological University, Houghton, MI, USA
\item[$^{84}$] New York University, New York, NY, USA
\item[$^{85}$] Northeastern University, Boston, MA, USA
\item[$^{86}$] Ohio State University, Columbus, OH, USA
\item[$^{87}$] Pennsylvania State University, University Park, PA, USA
\item[$^{88}$] University of Chicago, Enrico Fermi Institute, Chicago, IL, USA
\item[$^{89}$] University of Nebraska, Lincoln, NE, USA
\item[] -----
\item[$^{a}$] School of Physics and Astronomy, University of Leeds, Leeds, United Kingdom
\item[$^{b}$] Max-Planck-Institut f\"ur Radioastronomie, Bonn, Germany
\item[$^{c}$] also at Vrije Universiteit Brussels, Brussels, Belgium
\item[$^{d}$] now at Deutsches Elektronen-Synchrotron (DESY), Zeuthen, Germany
\item[$^{e}$] SUBATECH, \'Ecole des Mines de Nantes, CNRS-IN2P3, Universit\'e de Nantes, France
\item[$^{f}$] Fermi National Accelerator Laboratory, USA
\item[$^{g}$] University of Bucharest, Physics Department, Bucharest, Romania
\item[$^{h}$] Colorado State University, Fort Collins, CO
\item[$^{i}$] University of Hawaii, Honolulu, HI, USA
\item[$^{j}$] University of New Mexico, Albuquerque, NM, USA
\item[$^{k}$] now at Universit\'e Libre de Bruxelles (ULB), Brussels, Belgium
\end{description}